\documentclass[aps,prl,showpacs,twocolumn,floatfix,superscriptaddress]{revtex4-1}

\usepackage{graphicx}
\usepackage{amssymb,amsmath,bbm,braket,indentfirst,bm,setspace,amsthm,theorem,pifont}
\usepackage[colorlinks=true,linktoc=page,linkcolor=blue,citecolor=magenta]{hyperref}
\usepackage[normalem]{ulem}   
\usepackage{cancel,soul}
\usepackage[pdftex]{color}

\allowdisplaybreaks 

\makeatletter
\renewcommand*\env@matrix[1][*\c@MaxMatrixCols c]{%
  \hskip -\arraycolsep
  \let\@ifnextchar\new@ifnextchar
  \array{#1}}
\makeatother

\renewcommand\({\ensuremath \left(}
\renewcommand\){\ensuremath \right)}
\renewcommand\[{\ensuremath \left[}
\renewcommand\]{\ensuremath \right]}
\def\:={\,\raisebox{0.85pt}{.}\hspace{-2.78pt}\raisebox{2.85pt}{.}\!\!=\,}
\def\=:{\,=\!\!\raisebox{0.85pt}{.}\hspace{-2.78pt}\raisebox{2.85pt}{.}\,}

\newcommand{\tZ}{\widetilde{Z}}
\newcommand{\tU}{\widetilde{U}}

\begin{document}
\title{Floquet Supersymmetry}

\author{Thomas~Iadecola} \affiliation{Physics Department, Boston
University, Boston, Massachusetts 02215, USA}
\affiliation{Kavli Institute for Theoretical Physics, University of California,
Santa Barbara, California 93106, USA }
\affiliation{Joint Quantum Institute and Condensed Matter Theory Center,
Department of Physics, University of Maryland, College Park, Maryland 20742, USA}

\author{Timothy~H.~Hsieh}
\affiliation{Kavli Institute for Theoretical Physics, University of California,
Santa Barbara, California 93106, USA }
\affiliation{Perimeter Institute for Theoretical Physics,
Waterloo, Ontario N2L 2Y5, Canada}

\date{\today} 
\begin{abstract}
We show that time-reflection symmetry in periodically driven (Floquet) quantum systems enables an inherently nonequilibrium phenomenon structurally similar to quantum-mechanical sypersymmetry.  In particular, we find Floquet analogues of the Witten index that place lower bounds on the degeneracies of states with quasienergies $0$ and $\pi$.  Moreover, we show that in some cases time reflection symmetry can also interchange fermions and bosons, leading to fermion/boson pairs with opposite quasienergy. We provide a simple class of disordered, interacting, and ergodic Floquet models with an exponentially large number of states at quasienergies $0$ and $\pi$, which are robust as long as the time-reflection symmetry is preserved.  Floquet supersymmetry manifests itself in the evolution of certain local observables as a period-doubling effect with dramatic finite-size scaling, providing a clear signature for experiments.    
\end{abstract}
\maketitle

Quantum systems driven by time-periodic perturbations are ubiquitous in 
atomic, molecular, and optical physics~\cite{Shirley65,Sambe73,Bukov15a}.  In recent years, periodic driving
has been exploited by theory~\cite{Rahav03,Eckardt05,Oka09,Lindner11,Dalibard11,D'Alessio13,Goldman14,Grushin14,Iadecola15}
and experiment~\cite{Yin09,Aidelsburger13,Miyake13,Jotzu14,Bordia17}
as a resource for quantum simulation;
by varying certain control parameters periodically in time,
intricate effective Hamiltonians can be realized for synthetic quantum systems
that might be outlandish in the context of solid-state physics. However,
the analogy between static and periodically-driven (Floquet) quantum matter only goes so far.
Being time-dependent, Floquet systems
do not conserve energy and 
generically heat up to infinite temperature.  They lose any discernible phase structure~\cite{D'Alessio14,Lazarides14b} unless
some notion of integrability~\cite{Lazarides14a,Chandran16,Gritsev17}, many-body localization 
(MBL)~\cite{D'Alessio13,Huse13,Chandran14,Ponte15,Khemani16},
or prethermalization~\cite{Abanin15,Bukov15b,Mori16,Kuwahara16,Canovi16,Abanin17,Else17,Vajna17}
is invoked.

With any of these three stabilizing mechanisms, Floquet systems have exhibited many new phases that lack equilibrium counterparts.
These include a vast array of Floquet topological
phases~\cite{vonKeyserlingk16a,Else16a,Roy16,Potter16,Potirniche17} and the so-called
``$\pi$ spin glass" ($\pi$SG)~\cite{Khemani16,vonKeyserlingk16b}
or ``discrete time crystal" (DTC)~\cite{Else16b,Yao17} phase,
which are the objectives of recent
experiments~\cite{Zhang17,Choi17}. These
Floquet phases share qualitative features that stem from their nonequilibrium nature;
for example, in all such phases there are certain operators whose dynamics synchronizes robustly with the periodic
drive in a nontrivial way. This is especially striking in the $\pi$SG/DTC, where the
local magnetization exhibits robust subharmonic response at half the driving frequency.



In this work, we introduce a distinct class of Floquet systems that also exhibits subharmonic response, but for fundamentally different reasons than the $\pi$SG/DTC.  We dub this phenomenon ``Floquet supersymmetry'' (FSUSY), as the underlying structure has many close parallels to quantum-mechanical supersymmetry (SUSY).  For instance, while SUSY exchanges bosons and fermions, FSUSY exchanges forward and backward time evolution---its generator is a \textit{time-reflection symmetry}.  Interestingly, as we will show, in some cases the time reflection operator can {\it also} interchange bosons and fermions, leading to pairs of bosonic and fermionic states {\it at opposite quasienergies}. SUSY models are characterized by an invariant, the Witten index, which provides a lower bound on the ground-state degeneracy; similarly, FSUSY models are characterized by \textit{two} invariants, which place lower bounds on the degeneracies of the ``quasienergies" $0$ and $\pi$. We emphasize, however, that FSUSY is not simply a generalization of SUSY to the Floquet context; rather, as we will show, it is a distinct property of the time evolution operator of a Floquet system.

After establishing this general framework, we present a simple class of interacting, disordered, and \textit{ergodic} Floquet models exhibiting FSUSY.  In these models, the degeneracies of the $0,\pi$ quasienergies are exponentially large---at least $2^{L/2}$, where $L$ is the system size.  We show that this exponentially large
degeneracy is robust to any disorder and interactions preserving the underlying time-reflection symmetry.  Such models show a distinct experimental signature of FSUSY.  Local observables exhibit a subharmonic response; however, in stark contrast to the $\pi$SG/DTC, the response is suppressed exponentially in system size. This finite-size scaling of the response serves as sharp evidence of FSUSY.  It is remarkable that this subharmonic response occurs in an otherwise ergodic quantum system;
FSUSY provides an example of a class of thermalizing Floquet systems which display nontrivial phenomena in a macroscopic subspace of the full Hilbert space.  Nevertheless, there is no contradiction with ergodicity, as the subharmonic response scales to zero in the thermodynamic limit for generic initial states.

We begin with some definitions.  Consider a periodically driven system with the time-dependent
Hamiltonian $H(t+T)=H(t)$, with $T$ the driving period  
hereafter set to $1$ (along with $\hbar$).
Define the Floquet unitary
$U^{\,}_{\rm F}$, which evolves states by one period:
\begin{align}
\label{eq: U_F action}
U^{\,}_{\rm F}\ket{\psi(t)}=\ket{\psi(t+1)}.
\end{align}
$U^{\,}_{\rm F}$ has eigenstates \{$\ket{E}$\} with corresponding eigenvalues  \{$e^{\mathrm{i}E}$\}; the quasienergies \{$E$\} are defined modulo $2\pi$.

We say that $U^{\,}_{\rm F}$ has time-reflection symmetry if there exists a unitary operator
$R$ satisfying $R^{2}=\mathbbm{1}$ and 
\begin{align}
\label{eq: R definition}
R U^{\,}_{\rm F} R^{\dagger} = e^{\mathrm{i}\theta} U^{\dagger}_{\rm F}.
\end{align}
We hereafter set $\theta \rightarrow 0$ by redefining $U^{\,}_{\rm F} \rightarrow e^{\mathrm{i}\theta/2} U^{\,}_{\rm F}$.
Since $R$ maps the ``forward" Floquet evolution operator $U^{\,}_{\rm F}$
to the ``backward" Floquet evolution operator $U^{\dagger}_{\rm F}$, it can be interpreted as reversing
the direction of time. However, unlike the usual time-reversal operator, $R$ is
unitary, hence the name ``time-reflection symmetry." (The corresponding symmetry for the effective Hamiltonian is called chiral
symmetry, see e.g. Ref.~\cite{Asboth14}.)
Using Eq.~\eqref{eq: R definition},
we can deduce the action of $R$ on the Floquet eigenbasis:
\begin{align}
\label{eq: R action on U_F eigenbasis}
U^{\,}_{\rm F}\, (R \ket{E})
=
e^{-\mathrm{i}E}(R\ket{E}).
\end{align}
$R$ thus maps eigenstates of $U^{\,}_{\rm F}$ with quasienergy $E$ to
eigenstates of $U^{\,}_{\rm F}$ with quasienergy $-E$.  Hence, 
\begin{align}
\label{eq: ortho}
\bra{E}R\ket{E} = 0 \text{  if  } E\neq 0,\pi.
\end{align}
In the $E=0,\pi$ eigenspaces, $U^{\,}_{\rm F} = U^{\dagger}_{\rm F}$, so \eqref{eq: R definition} implies that $R$ and $U^{\,}_{\rm F}$ share a common eigenbasis for the $E=0,\pi$ states.  We will label the common eigenbasis for $E=0(\pi)$ as \{$\ket{0(\pi),\alpha}$\} where $\alpha=1,\dots,N^{\,}_{0(\pi)}$ and $N^{\,}_{0(\pi)}$ is the degeneracy of the $E=0(\pi)$ eigenspace.  Because $R^{2}=\mathbbm{1}$,
\begin{align}
\label{eq: ones}
\begin{split}
\bra{0,\alpha}R\ket{0,\alpha} &= \pm 1 \\
\bra{\pi,\alpha}R\ket{\pi,\alpha} &= \pm 1.
\end{split}
\end{align}

These properties motivate the definition of two trace formulas which we will prove to be integers providing lower bounds for the degeneracies $N^{\,}_{0,\pi}$.  Define $\mathcal{I}^{\,}_{0}, \mathcal{I}^{\,}_{\pi}$ as
\begin{align}
\label{eq: trace}
\begin{split}
\mathcal{I}^{\,}_{0 (\pi)}
&\equiv \text{tr}\(R\ \frac{U^{\,}_{\rm F}\pm\mathbbm{1}}{2}\) \\
&= \sum_{\alpha=1}^{N^{\,}_{0(\pi)}}
\bra{0(\pi),\alpha}R\ket{0(\pi),\alpha},
\end{split}
\end{align}
where we have used \eqref{eq: ortho}.  Moreover, \eqref{eq: ones} implies that both invariants are integers and $|\mathcal{I}^{\,}_{0(\pi)}| \leq N^{\,}_{0(\pi)}$; these invariants thus provide a lower bound for the number of $0$ and $\pi$ quasienergy eigenstates, respectively.

Given time reflection symmetry, $\mathcal{I}^{\,}_{0 (\pi)}$
are topological invariants in the following sense.  Consider any small perturbation to $U^{\,}_F$ which preserves the time-reflection symmetry \eqref{eq: R definition}.  We expect that the time reflection operator $R$ for which \eqref{eq: R definition} holds will change continuously as $U^{\,}_F$ is perturbed; we illustrate this later in a concrete model.  Since the trace is a continuous function of $R$ and $U^{\,}_F$, small changes in the arguments must lead to small changes in $\mathcal{I}^{\,}_{0 (\pi)}$. Because the latter are integers, they must remain invariant.  Hence, any symmetry-respecting perturbation continuously connected to the identity operator will not change the trace invariants $\mathcal{I}^{\,}_{0 (\pi)}$.  
We emphasize that the existence of these invariants and the subsequent properties depends essentially on the presence of time-reflection symmetry.

At this point, it is useful to draw contrasts and comparisons with ground states of static systems.  At first glance, this symmetry-protected ``pinning" of quasienergy eigenvalues to $0$ or $\pi$ may be reminiscent of
the protection of certain zero-energy modes in symmetry protected topological phases~\cite{JackiwRebbi,SSH,JackiwRossi,ReadGreen,TeoKane}.  In systems with topological defects or boundaries, zero modes may appear as bound states
protected by index theorems that define topological invariants similar to
Eq.~\eqref{eq: trace}~\cite{AtiyahSinger,Weinberg}.  However, in our Floquet setting there is no such bulk-boundary correspondence nor defects; the symmetry-protected $0$ and $\pi$ quasienergy modes are bulk entities. 
 
In fact, the closest static analogues of these protected many-body degeneracies arise in SUSY~\cite{Nicolai76,Witten81}, where the relevant topological invariant is the Witten index~\cite{Witten82} $\text{tr}[(-1)^F  e^{-\beta H}]$, with $H$ the Hamiltonian, $F$ the fermion number, and $\beta$ the inverse temperature.  The (integer) Witten index places a lower bound on the number of eigenstates at zero energy, and thereby on the ground-state degeneracy of $H$.  One remarkable phenomenon that can arise in certain SUSY models is ``superfrustration," where the Witten index scales exponentially with system size~\cite{Nicolai76,Fendley05,vanEerten05,Katsura17}.

Within this algebraic framework, there is potential for another connection to SUSY.  In systems with a conserved fermion parity $(-1)^F$, one possibility is that $\{(-1)^F,R\}=0$, in which case time-reflection changes the fermion parity of an eigenstate.  In this case, any bosonic state at quasienergy $E$ must have a fermionic partner with quasienergy $-E$.
This is in stark contrast to conventional SUSY, which exhibits pairs of bosonic and fermionic states at \textit{the same} energy.

We now present a simple model of FSUSY that features exponentially large degeneracy for even system sizes, and boson/fermion partners at equal and opposite quasienergy for odd system sizes. For the sake of exposition, we begin with the simplest model below and add interactions later. Consider a spin-$\frac{1}{2}$
chain with $L$ sites and the two part drive 
\begin{subequations}
\label{eq: model def}
\begin{align}
\label{eq: U_F def}
U_{\rm F}&=U^{\,}_{ZZ}\, U^{\,}_X, \\
U^{\,}_{ZZ} &\equiv \text{exp}\left(\mathrm{i}\frac{\pi}{4}\sum^{L}_{i=1}Z^{\,}_{i}Z^{\,}_{i+1}\right) \\
U^{\,}_{X}  &\equiv \text{exp}\left(-\mathrm{i}\sum^L_{i=1} h^{\,}_i X^{\,}_i\right).
\end{align}
\end{subequations}
Here, $X^{\,}_{i},Z^{\,}_{i}$ are Pauli operators on the site $i$, and $h^{\,}_i$ are random couplings.  We hereafter impose periodic boundary conditions (identifying sites $1$ and $L+1$).

The model \eqref{eq: model def} has time-reflection symmetry, generated by the operator
\begin{align}
\label{eq: R model def}
R^{\,}_{1}=U^{\dagger}_{X}\prod^{L}_{i=1}Z^{\,}_{i}.
\end{align}
To see that $R^{2}_{1}=\mathbbm{1}$, one can rewrite $R^{\,}_{1}=U^{1/2 \dagger}_{X}\,(\prod^{L}_{i=1}Z^{\,}_{i})\ U^{1/2}_{X}\equiv \prod^{L}_{i=1}\tZ^{\,}_{i}$, where $\tZ^{\,}_i \equiv e^{\mathrm{i}\,h^{\,}_i X^{\,}_i/2}\ Z^{\,}_i\ e^{-\mathrm{i}\,h^{\,}_i X^{\,}_i/2}$.
Using the fact that
$U^{\,}_{ZZ}=\mathrm{i}^L\, U^{\dagger}_{ZZ}$,
one verifies that Eq.~\eqref{eq: R definition} holds with $\theta=L\pi/2$. Observe that 
$R^{\,}_{1}$ depends explicitly on $U^{\,}_{\rm F}$, just as the generator of SUSY depends explicitly
on the Hamiltonian. 

In fact, one can define another time-reflection operator
\begin{align}
R^{\,}_{2} =U^{\,}_{ZZ} \prod^{L}_{i=1}Z^{\,}_{i}
\end{align}
with the requisite properties (setting $\theta=L\pi/2$), and $U^{\,}_F = R^{\,}_{2} R^{\,}_{1}$.  Again, this parallels SUSY, in which the Hamiltonian is constructed from the SUSY generators~\cite{Witten81}.  

Having established time-reflection symmetries in this model, we calculate the trace invariants \eqref{eq: trace} and find
\begin{align}
|\mathcal{I}^{\,}_{0 (\pi)}|
=
\begin{cases}
2^{L/2} & L\ \text{even}\\
0 & L\ \text{odd}
\end{cases}
\end{align}
(for both $R^{\,}_{1,2}$). Thus, for even system sizes, there is an exponentially large number of states with quasienergy $0,\pi$.
(See also~\cite{SM}.)

The above model can be rewritten in terms of
free fermions via a Jordan-Wigner transformation. 
Interestingly, the fermion parity operator $(-1)^F=\prod^L_{i=1}X^{\,}_{i}$
(anti)commutes with the time reflection operators for even (odd) $L$.  Thus, while the odd-$L$ case does not host an exponentially large $|\mathcal{I}^{\,}_{0 (\pi)}|$, it does exhibit an unconventional pairing of bosonic and fermionic states at equal and opposite quasienergy.

Crucially, the above properties are {\it not} artifacts of free fermions; the invariants $\mathcal{I}^{\,}_{0,\pi}$ are robust to any interaction that preserves time-reflection symmetry, while the pairing of bosonic and fermionic states additionally requires maintaining fermion parity conservation (Ising symmetry in the spin language).  To illustrate this, we add interactions to the transverse-field part of the drive: 
\begin{align}
\label{eq: H_int replacement}
U^{\,}_X \rightarrow U^{\,}_H \equiv \text{exp}\left[-\mathrm{i}\(\sum^L_{i=1} h^{\,}_i X^{\,}_i + g H^{\,}_{\rm int}\)\right],
\end{align}
where $g$ parameterizes the strength of interaction and we demand that $H^{\,}_{\rm int}$ anticommutes with $\prod^{L}_{i=1}Z^{\,}_{i}$.  The modified system then maintains time-reflection symmetry, with the modified time-reflection operator $R=U^{\dagger}_H\prod^{L}_{i=1}Z^{\,}_{i}$. (If $H^{\,}_{\rm int}$ additionally commutes with $\prod^L_{i=1}X^{\,}_{i}$, then the
unconventional boson/fermion pairing for odd $L$ also remains.)  As a result, the trace invariants (and exponentially large degeneracies) remain the same even in the presence of these interactions.  In~\cite{SM}, we provide an alternative way to understand the degeneracy as arising from the intersection of two large subspaces; this derivation also explains why the model's properties are robust to interactions of the above type.
  
We now focus on the case of even $L$, and investigate some consequences of the
macroscopic degeneracies protected by the indices $\mathcal{I}^{\,}_{0,\pi}$ For the purpose of numerics, we specify to the choice
\begin{align}
\label{eq: interactions}
\begin{split}
H^{\,}_{\rm int}
&=
\sum^{L}_{i=1}
\(
J^{xz}_{i} X^{\,}_{i}Z^{\,}_{i+1}
+J^{xxx}_{i} X^{\,}_{i-1}X^{\,}_{i}X^{\,}_{i+1}
\right. 
\\
&\qquad\qquad
\left.
+J^{zxz}_{i} Z^{\,}_{i-1}X^{\,}_{i}Z^{\,}_{i+1}\)
,
\end{split}
\end{align}
and we draw the random couplings $h^{\,}_{i},J^{xz}_{i},J^{xxx}_{i},$ and $J^{zxz}_{i}$ uniformly
from the interval $[0,\pi/2]$. 

\begin{figure}[t]
\includegraphics[width=0.475\textwidth]{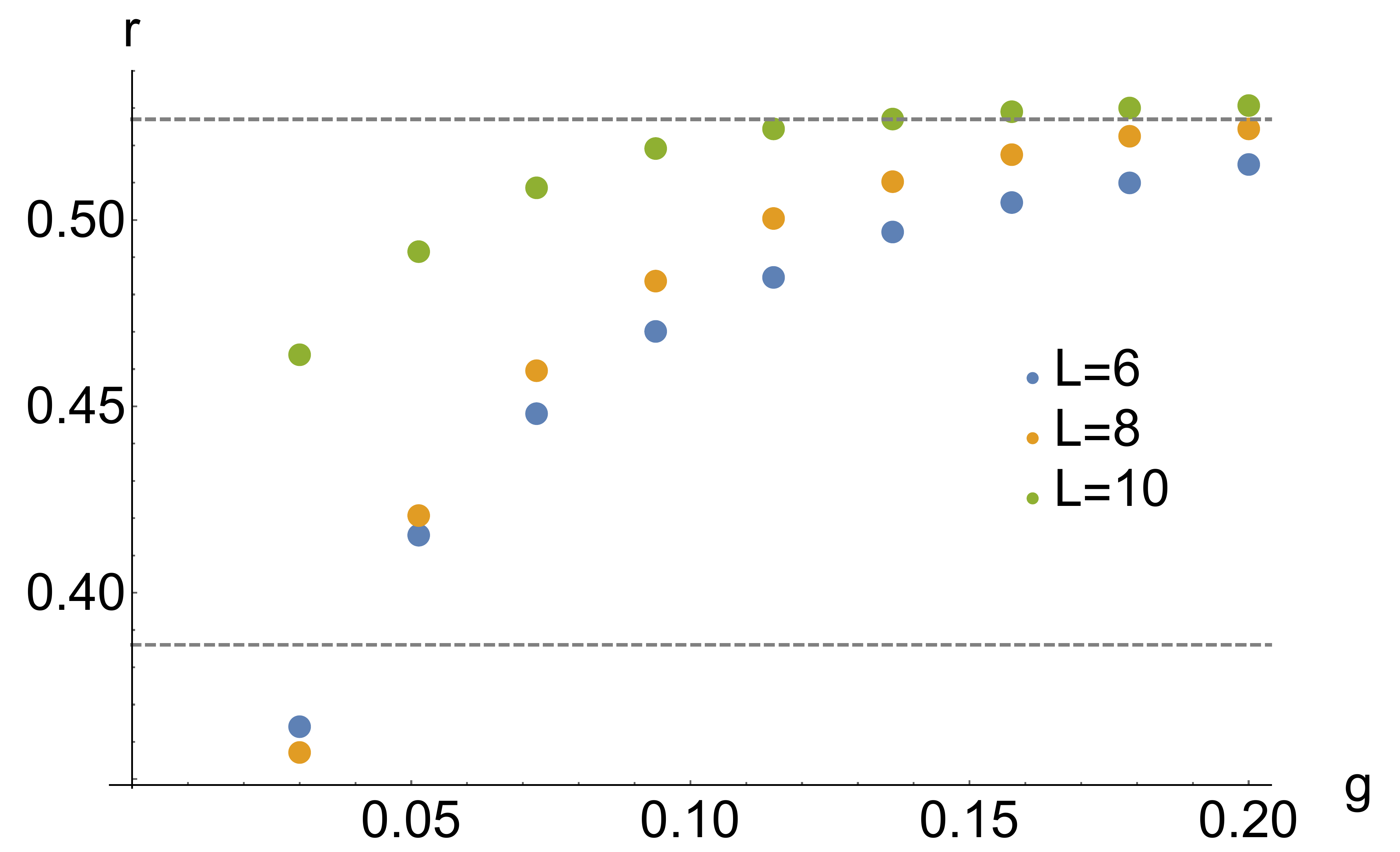}
\caption{
(Color online)
Statistics of quasienergy levels outside of the degenerate subspaces in the model \eqref{eq: model def}, as measured by the parameter $r$ defined in Ref.~\cite{Pal10}. 
All data points are averaged over
disorder realizations and quasienergy (see main text). Gray dashed lines at $r=0.386$ and $0.527$ indicate the
expected values for the Poisson and Wigner-Dyson distributions, respectively.
}
\label{fig: level statistics}
\end{figure}

\begin{figure}[t]
\includegraphics[width=0.475\textwidth]{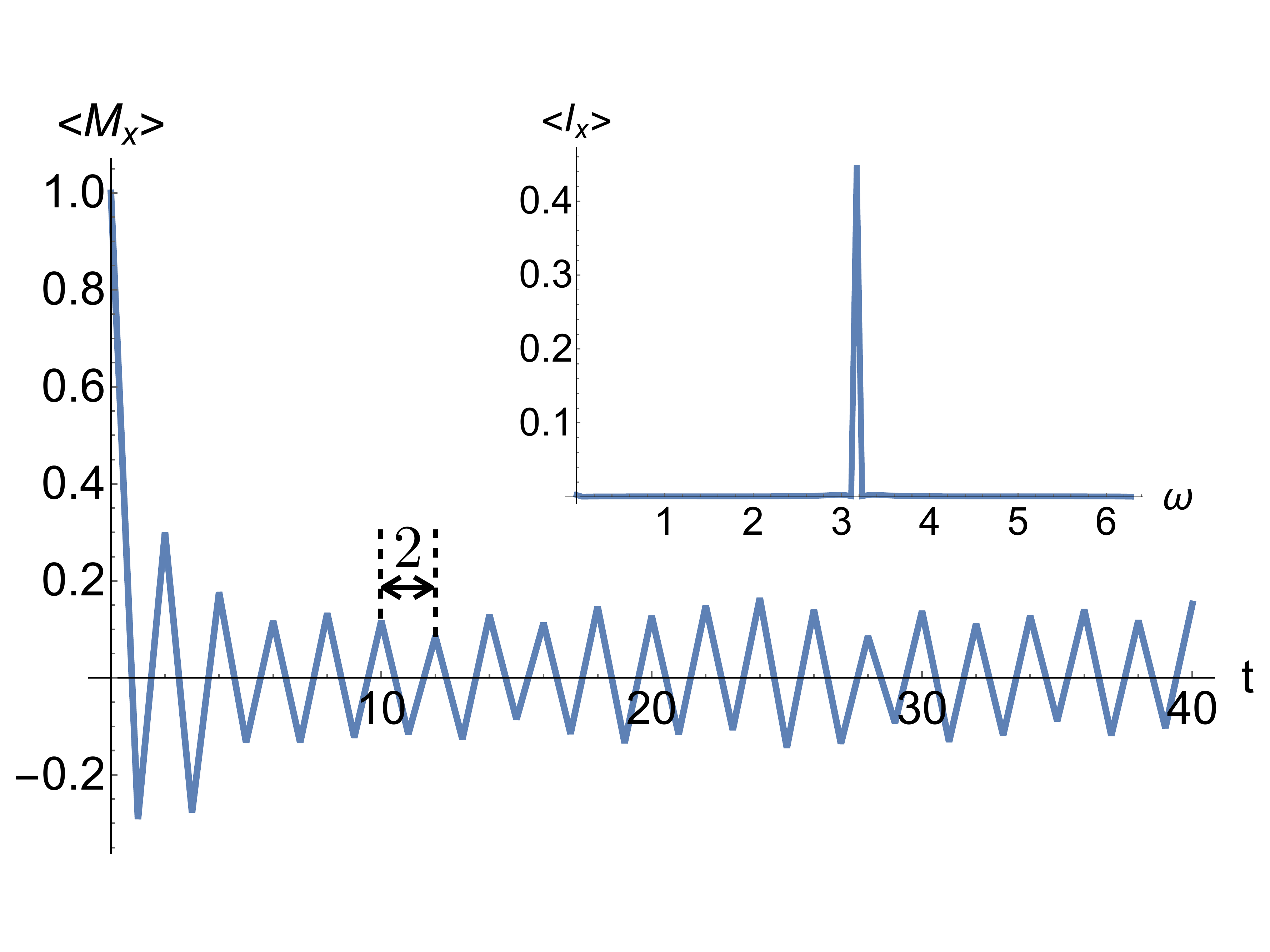}
\caption{
(Color online)
A representative time-series of the magnetization $M^{\,}_{X}$ starting from an initial state with all spins
polarized in the $X=+1$ direction, for a single disorder realization in the interacting version of the model
\eqref{eq: model def} with $L=8$ and $g=1$ [see Eqs.~\eqref{eq: H_int replacement} and \eqref{eq: interactions}].
Period-$2$ oscillations around the expected infinite-temperature
value $\langle M^{\,}_{X}\rangle=0$ are clearly visible at late times. Inset:
power spectrum $\langle I^{\,}_{X}(\omega)\rangle$ for the same $L$ and $g$, averaged over $20000$
disorder realizations.  The dominant coherent structure in the power spectrum is the peak at $\omega=\pi$,
which results from the period-$2$ oscillations.
        }
\label{fig: time series}
\end{figure}

One might wonder whether the symmetry constraint \eqref{eq: R definition}, which is evidently strong enough
to protect exponentially large degeneracies, is also strong enough to constrain the many-body spectrum outside of the degenerate subspaces.  Given the presence of strong disorder, are there signatures of many-body localization in this system?  To answer these questions,
we performed exact diagonalization at system sizes $L=6,8,$ and $10$ (for $20000$, $10000$, and
$5000$ disorder realizations, respectively) and computed the disorder-averaged
level statistics of the states outside the
$0,\pi$ subspaces.  We computed the parameter $r$ \cite{Pal10}; given an ordered list $\{E^{\,}_{j}\}$ of quasienergies, $r$ is defined in terms of the quasienergy gaps $\delta^{\,}_{j}\equiv E^{\,}_{j+1}-E^{\,}_{j}$ as the average
of the quantity $r^{\,}_j=\text{min}(\delta^{\,}_{j},\delta^{\,}_{j+1})/\text{max}(\delta^{\,}_{j},\delta^{\,}_{j+1})$ over quasienergy ($j$) and disorder realizations. Even at these
very small system sizes, we see level statistics consistent with the Wigner-Dyson distribution for
$g\gtrsim 0.2$ (see Fig.~\ref{fig: level statistics}). Thus, apart from the protected degeneracies, the model
\eqref{eq: model def} appears to be a generic ergodic system.

\begin{figure}[t]
\includegraphics[width=0.475\textwidth]{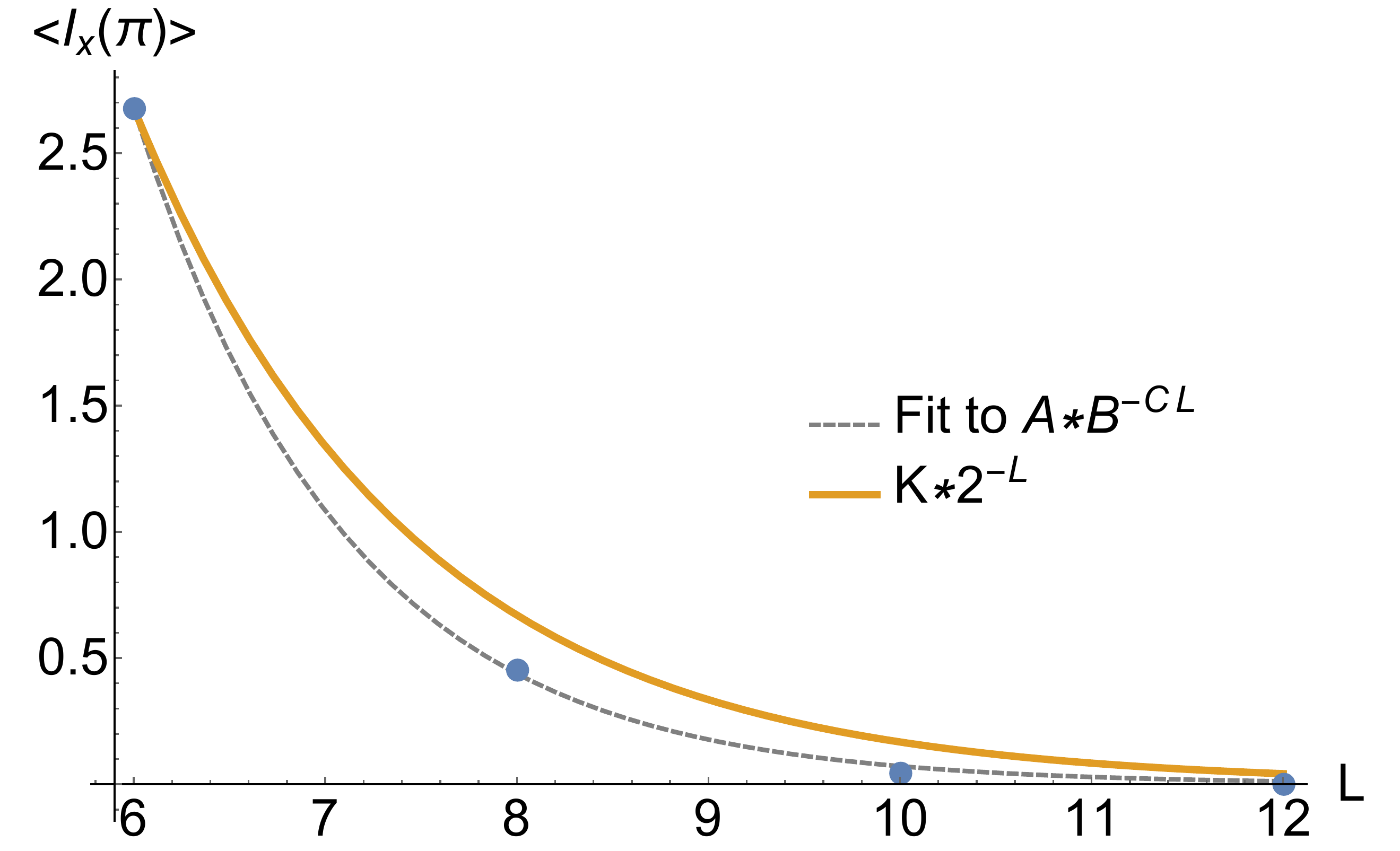}
\caption{
(Color online)
Magnitude of the peak in $\langle I^{\,}_{X}(\omega)\rangle$ at $\omega=\pi$ as a function
of system size, with an exponential fit (gray, dashed line) and the estimate \eqref{eq: estimate} (orange)
plotted for reference. The model used to generate the data is that used in Fig.~\ref{fig: time series}.  Data are averaged over $40000$, $20000$, $10000$, and $5000$ disorder realizations for
$L=6$, $8$, $10$, and $12$, respectively.
        }
\label{fig: finite size}
\end{figure}

Nonetheless, we now show that the protected degeneracies give rise to a distinct subharmonic response which serves as a direct signature of Floquet supersymmetry.  In particular, the time evolution of the expectation values of certain operators exhibit period-$2$ oscillations.  This follows directly from the existence of $0,\pi$ states, which are protected by FSUSY.  Assume there is at least one protected pair of states with quasienergy $0,\pi$, and denote by $D$ the space spanned by the two states.  Then the Floquet operator restricted to $D$ can be represented by $Z P^{\,}_D$, where $P^{\,}_D$ is the projection onto $D$ and $Z$ is a Pauli-$Z$ operator in the basis of the $0,\pi$ quasienergy states.  Hence, the operator $X P^{\,}_D$ will flip sign every period, as $\{Z,X\}=0$.  Note that in this general discussion $Z P^{\,}_D$ and $X P^{\,}_D$ may be nonlocal operators;  however, in the model \eqref{eq: model def}, there is a local operator, namely the on-site $X^{\,}_i$, which flips sign under the Floquet evolution restricted to the degenerate subspaces (see~\cite{SM}).

Therefore, in the time evolution of $X = X P^{\,}_D + X (1-P^{\,}_D)$, the latter piece will decay to zero because the complement of $D$ is generically ergodic, while the former piece contributes the period-$2$ oscillations.  However, the ratio of the size of $D$ to that of the entire Hilbert space decreases exponentially with system size $L$.  Hence, if one evolves from a random initial state, then the amplitude of such period-$2$ oscillations will decrease exponentially with $L$, a phenomenon that distinguishes FSUSY from the $\pi$SG/DTC phase.  In fact, such dependence on system size also occurs in signatures of SUSY in Majorana models with translation symmetry \cite{Hsieh16}.     

We have checked these properties in the above model by computing the time evolution of the total magnetization
$M^{\,}_{X}=\frac{1}{L}\sum^{L}_{i=1}X^{\,}_{i}$ starting from an initial state with all spins polarized in the 
$X=+1$ direction.  A representative time series for $L=8$ is shown in Fig.~\ref{fig: time series}. Plots of the
expectation values of single-site $X^{\,}_{i}$ operators look similar.  A useful figure of merit for quantifying this
subharmonic response is the power spectrum $\langle I^{\,}_{X}(\omega)\rangle$, obtained by taking the
modulus-squared of the Fourier transform of $\langle M^{\,}_{X}(t)\rangle$, which displays a peak at $\omega=\pi$
if $\langle M^{\,}_{X}(t)\rangle$ exhibits period-$2$ oscillations. We indeed find such behavior in the power
spectrum; averaging over disorder realizations, we find a single peak at $\omega=\pi$, and all other structure
washes out (see Fig.~\ref{fig: time series} inset).

For a typical initial state, which has overlap with all eigenstates
of $U^{\,}_{\rm F}$, we can estimate (up to a multiplicative prefactor) that
\begin{align}
\label{eq: estimate}
\langle I^{\,}_{\!X}\!(\pi)\rangle
&\!=\!
\Big\vert
\!
\sum_{\substack{E,E^{\prime}\!, \\ \alpha,\alpha^{\prime}}}
\!c^*_{E, \alpha}c^{\,}_{E^{\prime}, \alpha^{\prime}}
\!\bra{E,\!\alpha}
\!M_X\!
\ket{E^{\prime}\!,\!\alpha^{\prime}}
\delta(E\!-\!E^{\prime}\!-\!\pi)
\Big\vert^2
\!\! \nonumber
\\
&\lesssim
2^{-L}, 
\end{align}
where $c^{\,}_{E,\alpha}=\braket{E,\alpha|\psi}$ is the overlap of eigenstates with the initial state $\ket{\psi}$.
This exponential upper bound on the finite-size scaling of $\langle I^{\,}_{X}(\pi)\rangle$ results from
the fact that the degenerate quasienergy eigenstates constitute a fraction of order $2^{-L/2}$ of all eigenstates
of $U^{\,}_{\rm F}$.  We see finite-size scaling of the disorder-average of $\langle I^{\,}_{X}(\pi)\rangle$
in exact diagonalization that is consistent with this estimate (see Fig.~\ref{fig: finite size}). Our simulations were
carried out at $g=1$, so that the energy levels outside the degenerate subspaces are approximately
Wigner-Dyson-distributed. It is interesting that even in this chaotic regime, there are still coherent period-$2$ oscillations. Although this effect
disappears in the thermodynamic limit due to the exponential suppression described above, it should be accessible in quantum
simulation experiments, which are performed at a variable finite size.

We note that the persistence of the oscillations described above depend crucially on the presence of time-reflection symmetry; without it, the oscillations acquire a finite lifetime.  However, in~\cite{SM} we show that, for sufficiently small time-reflection breaking, the oscillations can persist long enough to be experimentally observable.

There are several interesting avenues to pursue regarding both Floquet supersymmetry and the particular class of models presented.  FSUSY provides an alternative mechanism for achieving subharmonic response; whereas the robustness in the discrete time crystal relies on the rigidity of eigenstates (long-range correlations in space), the robustness in FSUSY relies on the rigidity of the eigenvalues pinned to $0,\pi$, a consequence of the underlying time-reflection symmetry.  Moreover, FSUSY provides a mechanism whereby a protected subspace can exhibit nontrivial phenomena (e.g., period-$2$ oscillations) despite being embedded in a thermal system.  Thus, even though non-integrable systems without many-body localization may heat to infinite temperature, it may be possible that a subspace (whose dimension can grow exponentially with system size) can behave nontrivially, as FSUSY illustrates.     

The most pressing
question concerning the model \eqref{eq: model def} at even $L$ is that of the nature of the degenerate states---aside from their fixed quasienergy, do
they have any special properties that are not shared by the rest of the eigenstates of $U^{\,}_{\rm F}$? The derivation of the degenerate states as the intersection of two large subspaces in \cite{SM} suggests that the degenerate states may be highly entangled, but it would be useful to quantify the amount of entanglement.  It would also be interesting
to consider whether the protected macroscopic degeneracy could be useful for quantum information
processing.  Having access to an exponentially large number of exactly degenerate eigenstates
could aid in the coherent storage and manipulation of quantum information.

We especially thank C.-M.~Jian for useful discussions, as well as A. Chandran and V. Khemani.  T.I. gratefully acknowledges the hospitality of the KITP during the course of this work.  T.I.~was supported by the National Science Foundation Graduate Research Fellowship Program under Grant No.~DGE-1247312, the DOE under Grant No.~DE-FG02-06ER46316, the Laboratory for Physical Sciences, Microsoft, and a JQI Postdoctoral Fellowship.  T.H.H. was supported by the Gordon and Betty Moore Foundation's EPiQS Initiative through Grant No.~GBMF4304.  Research at Perimeter Institute is supported by the Government of
Canada through Industry Canada and by the Province of Ontario through the Ministry of Research and Innovation.

\bibliographystyle{apsrev}

\bibliography{refs_Floquet_degeneracy_paper}

\begin{thebibliography}{61}
\expandafter\ifx\csname natexlab\endcsname\relax\def\natexlab#1{#1}\fi
\expandafter\ifx\csname bibnamefont\endcsname\relax
  \def\bibnamefont#1{#1}\fi
\expandafter\ifx\csname bibfnamefont\endcsname\relax
  \def\bibfnamefont#1{#1}\fi
\expandafter\ifx\csname citenamefont\endcsname\relax
  \def\citenamefont#1{#1}\fi
\expandafter\ifx\csname url\endcsname\relax
  \def\url#1{\texttt{#1}}\fi
\expandafter\ifx\csname urlprefix\endcsname\relax\def\urlprefix{URL }\fi
\providecommand{\bibinfo}[2]{#2}
\providecommand{\eprint}[2][]{\url{#2}}

\bibitem[{\citenamefont{Shirley}(1965)}]{Shirley65}
\bibinfo{author}{\bibfnamefont{J.~H.} \bibnamefont{Shirley}},
  \bibinfo{journal}{Phys. Rev.} \textbf{\bibinfo{volume}{138}},
  \bibinfo{pages}{B979} (\bibinfo{year}{1965}).

\bibitem[{\citenamefont{Sambe}(1973)}]{Sambe73}
\bibinfo{author}{\bibfnamefont{H.}~\bibnamefont{Sambe}},
  \bibinfo{journal}{Phys. Rev. A} \textbf{\bibinfo{volume}{7}},
  \bibinfo{pages}{2203} (\bibinfo{year}{1973}).

\bibitem[{\citenamefont{Bukov et~al.}(2015{\natexlab{a}})\citenamefont{Bukov,
  D'Alessio, and Polkovnikov}}]{Bukov15a}
\bibinfo{author}{\bibfnamefont{M.}~\bibnamefont{Bukov}},
  \bibinfo{author}{\bibfnamefont{L.}~\bibnamefont{D'Alessio}},
  \bibnamefont{and}
  \bibinfo{author}{\bibfnamefont{A.}~\bibnamefont{Polkovnikov}},
  \bibinfo{journal}{Advances in Physics} \textbf{\bibinfo{volume}{64}},
  \bibinfo{pages}{139} (\bibinfo{year}{2015}{\natexlab{a}}).

\bibitem[{\citenamefont{Rahav et~al.}(2003)\citenamefont{Rahav, Gilary, and
  Fishman}}]{Rahav03}
\bibinfo{author}{\bibfnamefont{S.}~\bibnamefont{Rahav}},
  \bibinfo{author}{\bibfnamefont{I.}~\bibnamefont{Gilary}}, \bibnamefont{and}
  \bibinfo{author}{\bibfnamefont{S.}~\bibnamefont{Fishman}},
  \bibinfo{journal}{Phys. Rev. A} \textbf{\bibinfo{volume}{68}},
  \bibinfo{pages}{013820} (\bibinfo{year}{2003}).

\bibitem[{\citenamefont{Eckardt et~al.}(2005)\citenamefont{Eckardt, Weiss, and
  Holthaus}}]{Eckardt05}
\bibinfo{author}{\bibfnamefont{A.}~\bibnamefont{Eckardt}},
  \bibinfo{author}{\bibfnamefont{C.}~\bibnamefont{Weiss}}, \bibnamefont{and}
  \bibinfo{author}{\bibfnamefont{M.}~\bibnamefont{Holthaus}},
  \bibinfo{journal}{Phys. Rev. Lett.} \textbf{\bibinfo{volume}{95}},
  \bibinfo{pages}{260404} (\bibinfo{year}{2005}).

\bibitem[{\citenamefont{Oka and Aoki}(2009)}]{Oka09}
\bibinfo{author}{\bibfnamefont{T.}~\bibnamefont{Oka}} \bibnamefont{and}
  \bibinfo{author}{\bibfnamefont{H.}~\bibnamefont{Aoki}},
  \bibinfo{journal}{Phys. Rev. B} \textbf{\bibinfo{volume}{79}},
  \bibinfo{pages}{081406} (\bibinfo{year}{2009}).

\bibitem[{\citenamefont{Lindner et~al.}(2011)\citenamefont{Lindner, Refael, and
  Galitski}}]{Lindner11}
\bibinfo{author}{\bibfnamefont{N.~H.} \bibnamefont{Lindner}},
  \bibinfo{author}{\bibfnamefont{G.}~\bibnamefont{Refael}}, \bibnamefont{and}
  \bibinfo{author}{\bibfnamefont{V.}~\bibnamefont{Galitski}},
  \bibinfo{journal}{Nature Phys.} \textbf{\bibinfo{volume}{7}},
  \bibinfo{pages}{490} (\bibinfo{year}{2011}).

\bibitem[{\citenamefont{Dalibard et~al.}(2011)\citenamefont{Dalibard, Gerbier,
  Juzeliunas, and \"Ohberg}}]{Dalibard11}
\bibinfo{author}{\bibfnamefont{J.}~\bibnamefont{Dalibard}},
  \bibinfo{author}{\bibfnamefont{F.}~\bibnamefont{Gerbier}},
  \bibinfo{author}{\bibfnamefont{G.}~\bibnamefont{Juzeliunas}},
  \bibnamefont{and} \bibinfo{author}{\bibfnamefont{P.}~\bibnamefont{\"Ohberg}},
  \bibinfo{journal}{Rev. Mod. Phys.} \textbf{\bibinfo{volume}{83}},
  \bibinfo{pages}{1523} (\bibinfo{year}{2011}).

\bibitem[{\citenamefont{D'Alessio and Polkovnikov}(2013)}]{D'Alessio13}
\bibinfo{author}{\bibfnamefont{L.}~\bibnamefont{D'Alessio}} \bibnamefont{and}
  \bibinfo{author}{\bibfnamefont{A.}~\bibnamefont{Polkovnikov}},
  \bibinfo{journal}{Ann. Phys.} \textbf{\bibinfo{volume}{333}},
  \bibinfo{pages}{19 } (\bibinfo{year}{2013}).

\bibitem[{\citenamefont{Goldman and Dalibard}(2014)}]{Goldman14}
\bibinfo{author}{\bibfnamefont{N.}~\bibnamefont{Goldman}} \bibnamefont{and}
  \bibinfo{author}{\bibfnamefont{J.}~\bibnamefont{Dalibard}},
  \bibinfo{journal}{Phys. Rev. X} \textbf{\bibinfo{volume}{4}},
  \bibinfo{pages}{031027} (\bibinfo{year}{2014}).

\bibitem[{\citenamefont{Grushin et~al.}(2014)\citenamefont{Grushin,
  G\'omez-Le\'on, and Neupert}}]{Grushin14}
\bibinfo{author}{\bibfnamefont{A.~G.} \bibnamefont{Grushin}},
  \bibinfo{author}{\bibfnamefont{A.}~\bibnamefont{G\'omez-Le\'on}},
  \bibnamefont{and} \bibinfo{author}{\bibfnamefont{T.}~\bibnamefont{Neupert}},
  \bibinfo{journal}{Phys. Rev. Lett.} \textbf{\bibinfo{volume}{112}},
  \bibinfo{pages}{156801} (\bibinfo{year}{2014}).

\bibitem[{\citenamefont{Iadecola et~al.}(2015)\citenamefont{Iadecola, Santos,
  and Chamon}}]{Iadecola15}
\bibinfo{author}{\bibfnamefont{T.}~\bibnamefont{Iadecola}},
  \bibinfo{author}{\bibfnamefont{L.~H.} \bibnamefont{Santos}},
  \bibnamefont{and} \bibinfo{author}{\bibfnamefont{C.}~\bibnamefont{Chamon}},
  \bibinfo{journal}{Phys. Rev. B} \textbf{\bibinfo{volume}{92}},
  \bibinfo{pages}{125107} (\bibinfo{year}{2015}).

\bibitem[{\citenamefont{Lin et~al.}(2009)\citenamefont{Lin, Compton, Perry,
  Phillips, Porto, and Spielman}}]{Yin09}
\bibinfo{author}{\bibfnamefont{Y.-J.} \bibnamefont{Lin}},
  \bibinfo{author}{\bibfnamefont{R.~L.} \bibnamefont{Compton}},
  \bibinfo{author}{\bibfnamefont{A.~R.} \bibnamefont{Perry}},
  \bibinfo{author}{\bibfnamefont{W.~D.} \bibnamefont{Phillips}},
  \bibinfo{author}{\bibfnamefont{J.~V.} \bibnamefont{Porto}}, \bibnamefont{and}
  \bibinfo{author}{\bibfnamefont{I.~B.} \bibnamefont{Spielman}},
  \bibinfo{journal}{Phys. Rev. Lett.} \textbf{\bibinfo{volume}{102}},
  \bibinfo{pages}{130401} (\bibinfo{year}{2009}).

\bibitem[{\citenamefont{Aidelsburger et~al.}(2013)\citenamefont{Aidelsburger,
  Atala, Lohse, Barreiro, Paredes, and Bloch}}]{Aidelsburger13}
\bibinfo{author}{\bibfnamefont{M.}~\bibnamefont{Aidelsburger}},
  \bibinfo{author}{\bibfnamefont{M.}~\bibnamefont{Atala}},
  \bibinfo{author}{\bibfnamefont{M.}~\bibnamefont{Lohse}},
  \bibinfo{author}{\bibfnamefont{J.}~\bibnamefont{Barreiro}},
  \bibinfo{author}{\bibfnamefont{B.}~\bibnamefont{Paredes}}, \bibnamefont{and}
  \bibinfo{author}{\bibfnamefont{I.}~\bibnamefont{Bloch}},
  \bibinfo{journal}{Phys. Rev. Lett.} \textbf{\bibinfo{volume}{111}},
  \bibinfo{pages}{185301} (\bibinfo{year}{2013}).

\bibitem[{\citenamefont{Miyake et~al.}(2013)\citenamefont{Miyake, Siviloglou,
  Kennedy, Burton, and Ketterle}}]{Miyake13}
\bibinfo{author}{\bibfnamefont{H.}~\bibnamefont{Miyake}},
  \bibinfo{author}{\bibfnamefont{G.~A.} \bibnamefont{Siviloglou}},
  \bibinfo{author}{\bibfnamefont{C.~J.} \bibnamefont{Kennedy}},
  \bibinfo{author}{\bibfnamefont{W.~C.} \bibnamefont{Burton}},
  \bibnamefont{and} \bibinfo{author}{\bibfnamefont{W.}~\bibnamefont{Ketterle}},
  \bibinfo{journal}{Phys. Rev. Lett.} \textbf{\bibinfo{volume}{111}},
  \bibinfo{pages}{185302} (\bibinfo{year}{2013}).

\bibitem[{\citenamefont{Jotzu et~al.}(2014)\citenamefont{Jotzu, Messer,
  Desbuquois, Lebrat, Uehlinger, Greif, and Esslinger}}]{Jotzu14}
\bibinfo{author}{\bibfnamefont{G.}~\bibnamefont{Jotzu}},
  \bibinfo{author}{\bibfnamefont{M.}~\bibnamefont{Messer}},
  \bibinfo{author}{\bibfnamefont{R.}~\bibnamefont{Desbuquois}},
  \bibinfo{author}{\bibfnamefont{M.}~\bibnamefont{Lebrat}},
  \bibinfo{author}{\bibfnamefont{T.}~\bibnamefont{Uehlinger}},
  \bibinfo{author}{\bibfnamefont{D.}~\bibnamefont{Greif}}, \bibnamefont{and}
  \bibinfo{author}{\bibfnamefont{T.}~\bibnamefont{Esslinger}},
  \bibinfo{journal}{Nature} \textbf{\bibinfo{volume}{515}},
  \bibinfo{pages}{237} (\bibinfo{year}{2014}).

\bibitem[{\citenamefont{Bordia et~al.}(2017)\citenamefont{Bordia, L{\"u}schen,
  Schneider, Knap, and Bloch}}]{Bordia17}
\bibinfo{author}{\bibfnamefont{P.}~\bibnamefont{Bordia}},
  \bibinfo{author}{\bibfnamefont{H.}~\bibnamefont{L{\"u}schen}},
  \bibinfo{author}{\bibfnamefont{U.}~\bibnamefont{Schneider}},
  \bibinfo{author}{\bibfnamefont{M.}~\bibnamefont{Knap}}, \bibnamefont{and}
  \bibinfo{author}{\bibfnamefont{I.}~\bibnamefont{Bloch}},
  \bibinfo{journal}{Nat. Phys.} \textbf{\bibinfo{volume}{13}},
  \bibinfo{pages}{460} (\bibinfo{year}{2017}).

\bibitem[{\citenamefont{D'Alessio and Rigol}(2014)}]{D'Alessio14}
\bibinfo{author}{\bibfnamefont{L.}~\bibnamefont{D'Alessio}} \bibnamefont{and}
  \bibinfo{author}{\bibfnamefont{M.}~\bibnamefont{Rigol}},
  \bibinfo{journal}{Phys. Rev. X} \textbf{\bibinfo{volume}{4}},
  \bibinfo{pages}{041048} (\bibinfo{year}{2014}).

\bibitem[{\citenamefont{Lazarides
  et~al.}(2014{\natexlab{a}})\citenamefont{Lazarides, Das, and
  Moessner}}]{Lazarides14b}
\bibinfo{author}{\bibfnamefont{A.}~\bibnamefont{Lazarides}},
  \bibinfo{author}{\bibfnamefont{A.}~\bibnamefont{Das}}, \bibnamefont{and}
  \bibinfo{author}{\bibfnamefont{R.}~\bibnamefont{Moessner}},
  \bibinfo{journal}{Phys. Rev. E} \textbf{\bibinfo{volume}{90}},
  \bibinfo{pages}{012110} (\bibinfo{year}{2014}{\natexlab{a}}).

\bibitem[{\citenamefont{Lazarides
  et~al.}(2014{\natexlab{b}})\citenamefont{Lazarides, Das, and
  Moessner}}]{Lazarides14a}
\bibinfo{author}{\bibfnamefont{A.}~\bibnamefont{Lazarides}},
  \bibinfo{author}{\bibfnamefont{A.}~\bibnamefont{Das}}, \bibnamefont{and}
  \bibinfo{author}{\bibfnamefont{R.}~\bibnamefont{Moessner}},
  \bibinfo{journal}{Phys. Rev. Lett.} \textbf{\bibinfo{volume}{112}},
  \bibinfo{pages}{150401} (\bibinfo{year}{2014}{\natexlab{b}}).

\bibitem[{\citenamefont{Chandran and Sondhi}(2016)}]{Chandran16}
\bibinfo{author}{\bibfnamefont{A.}~\bibnamefont{Chandran}} \bibnamefont{and}
  \bibinfo{author}{\bibfnamefont{S.~L.} \bibnamefont{Sondhi}},
  \bibinfo{journal}{Phys. Rev. B} \textbf{\bibinfo{volume}{93}},
  \bibinfo{pages}{174305} (\bibinfo{year}{2016}).

\bibitem[{\citenamefont{Gritsev and Polkovnikov}(2017)}]{Gritsev17}
\bibinfo{author}{\bibfnamefont{V.}~\bibnamefont{Gritsev}} \bibnamefont{and}
  \bibinfo{author}{\bibfnamefont{A.}~\bibnamefont{Polkovnikov}},
  \bibinfo{journal}{SciPost Phys.} \textbf{\bibinfo{volume}{2}},
  \bibinfo{pages}{021} (\bibinfo{year}{2017}).

\bibitem[{\citenamefont{Huse et~al.}(2013)\citenamefont{Huse, Nandkishore,
  Oganesyan, Pal, and Sondhi}}]{Huse13}
\bibinfo{author}{\bibfnamefont{D.~A.} \bibnamefont{Huse}},
  \bibinfo{author}{\bibfnamefont{R.}~\bibnamefont{Nandkishore}},
  \bibinfo{author}{\bibfnamefont{V.}~\bibnamefont{Oganesyan}},
  \bibinfo{author}{\bibfnamefont{A.}~\bibnamefont{Pal}}, \bibnamefont{and}
  \bibinfo{author}{\bibfnamefont{S.~L.} \bibnamefont{Sondhi}},
  \bibinfo{journal}{Phys. Rev. B} \textbf{\bibinfo{volume}{88}},
  \bibinfo{pages}{014206} (\bibinfo{year}{2013}).

\bibitem[{\citenamefont{Chandran et~al.}(2014)\citenamefont{Chandran, Khemani,
  Laumann, and Sondhi}}]{Chandran14}
\bibinfo{author}{\bibfnamefont{A.}~\bibnamefont{Chandran}},
  \bibinfo{author}{\bibfnamefont{V.}~\bibnamefont{Khemani}},
  \bibinfo{author}{\bibfnamefont{C.~R.} \bibnamefont{Laumann}},
  \bibnamefont{and} \bibinfo{author}{\bibfnamefont{S.~L.}
  \bibnamefont{Sondhi}}, \bibinfo{journal}{Phys. Rev. B}
  \textbf{\bibinfo{volume}{89}}, \bibinfo{pages}{144201}
  (\bibinfo{year}{2014}).

\bibitem[{\citenamefont{Ponte et~al.}(2015)\citenamefont{Ponte, Chandran,
  Papi\'c, and Abanin}}]{Ponte15}
\bibinfo{author}{\bibfnamefont{P.}~\bibnamefont{Ponte}},
  \bibinfo{author}{\bibfnamefont{A.}~\bibnamefont{Chandran}},
  \bibinfo{author}{\bibfnamefont{Z.}~\bibnamefont{Papi\'c}}, \bibnamefont{and}
  \bibinfo{author}{\bibfnamefont{D.~A.} \bibnamefont{Abanin}},
  \bibinfo{journal}{Ann. Phys.} \textbf{\bibinfo{volume}{353}},
  \bibinfo{pages}{196 } (\bibinfo{year}{2015}).

\bibitem[{\citenamefont{Khemani et~al.}(2016)\citenamefont{Khemani, Lazarides,
  Moessner, and Sondhi}}]{Khemani16}
\bibinfo{author}{\bibfnamefont{V.}~\bibnamefont{Khemani}},
  \bibinfo{author}{\bibfnamefont{A.}~\bibnamefont{Lazarides}},
  \bibinfo{author}{\bibfnamefont{R.}~\bibnamefont{Moessner}}, \bibnamefont{and}
  \bibinfo{author}{\bibfnamefont{S.~L.} \bibnamefont{Sondhi}},
  \bibinfo{journal}{Phys. Rev. Lett.} \textbf{\bibinfo{volume}{116}},
  \bibinfo{pages}{250401} (\bibinfo{year}{2016}).

\bibitem[{\citenamefont{Abanin et~al.}(2015)\citenamefont{Abanin, De~Roeck, and
  Huveneers}}]{Abanin15}
\bibinfo{author}{\bibfnamefont{D.~A.} \bibnamefont{Abanin}},
  \bibinfo{author}{\bibfnamefont{W.}~\bibnamefont{De~Roeck}}, \bibnamefont{and}
  \bibinfo{author}{\bibfnamefont{F.}~\bibnamefont{Huveneers}},
  \bibinfo{journal}{Phys. Rev. Lett.} \textbf{\bibinfo{volume}{115}},
  \bibinfo{pages}{256803} (\bibinfo{year}{2015}).

\bibitem[{\citenamefont{Bukov et~al.}(2015{\natexlab{b}})\citenamefont{Bukov,
  Gopalakrishnan, Knap, and Demler}}]{Bukov15b}
\bibinfo{author}{\bibfnamefont{M.}~\bibnamefont{Bukov}},
  \bibinfo{author}{\bibfnamefont{S.}~\bibnamefont{Gopalakrishnan}},
  \bibinfo{author}{\bibfnamefont{M.}~\bibnamefont{Knap}}, \bibnamefont{and}
  \bibinfo{author}{\bibfnamefont{E.}~\bibnamefont{Demler}},
  \bibinfo{journal}{Phys. Rev. Lett.} \textbf{\bibinfo{volume}{115}},
  \bibinfo{pages}{205301} (\bibinfo{year}{2015}{\natexlab{b}}).

\bibitem[{\citenamefont{Mori et~al.}(2016)\citenamefont{Mori, Kuwahara, and
  Saito}}]{Mori16}
\bibinfo{author}{\bibfnamefont{T.}~\bibnamefont{Mori}},
  \bibinfo{author}{\bibfnamefont{T.}~\bibnamefont{Kuwahara}}, \bibnamefont{and}
  \bibinfo{author}{\bibfnamefont{K.}~\bibnamefont{Saito}},
  \bibinfo{journal}{Phys. Rev. Lett.} \textbf{\bibinfo{volume}{116}},
  \bibinfo{pages}{120401} (\bibinfo{year}{2016}).

\bibitem[{\citenamefont{Kuwahara et~al.}(2016)\citenamefont{Kuwahara, Mori, and
  Saito}}]{Kuwahara16}
\bibinfo{author}{\bibfnamefont{T.}~\bibnamefont{Kuwahara}},
  \bibinfo{author}{\bibfnamefont{T.}~\bibnamefont{Mori}}, \bibnamefont{and}
  \bibinfo{author}{\bibfnamefont{K.}~\bibnamefont{Saito}},
  \bibinfo{journal}{Ann. Phys.} \textbf{\bibinfo{volume}{367}},
  \bibinfo{pages}{96 } (\bibinfo{year}{2016}).

\bibitem[{\citenamefont{Canovi et~al.}(2016)\citenamefont{Canovi, Kollar, and
  Eckstein}}]{Canovi16}
\bibinfo{author}{\bibfnamefont{E.}~\bibnamefont{Canovi}},
  \bibinfo{author}{\bibfnamefont{M.}~\bibnamefont{Kollar}}, \bibnamefont{and}
  \bibinfo{author}{\bibfnamefont{M.}~\bibnamefont{Eckstein}},
  \bibinfo{journal}{Phys. Rev. E} \textbf{\bibinfo{volume}{93}},
  \bibinfo{pages}{012130} (\bibinfo{year}{2016}).

\bibitem[{\citenamefont{Abanin et~al.}(2017)\citenamefont{Abanin, De~Roeck, Ho,
  and Huveneers}}]{Abanin17}
\bibinfo{author}{\bibfnamefont{D.~A.} \bibnamefont{Abanin}},
  \bibinfo{author}{\bibfnamefont{W.}~\bibnamefont{De~Roeck}},
  \bibinfo{author}{\bibfnamefont{W.~W.} \bibnamefont{Ho}}, \bibnamefont{and}
  \bibinfo{author}{\bibfnamefont{F.}~\bibnamefont{Huveneers}},
  \bibinfo{journal}{Phys. Rev. B} \textbf{\bibinfo{volume}{95}},
  \bibinfo{pages}{014112} (\bibinfo{year}{2017}).

\bibitem[{\citenamefont{Else et~al.}(2017)\citenamefont{Else, Bauer, and
  Nayak}}]{Else17}
\bibinfo{author}{\bibfnamefont{D.~V.} \bibnamefont{Else}},
  \bibinfo{author}{\bibfnamefont{B.}~\bibnamefont{Bauer}}, \bibnamefont{and}
  \bibinfo{author}{\bibfnamefont{C.}~\bibnamefont{Nayak}},
  \bibinfo{journal}{Phys. Rev. X} \textbf{\bibinfo{volume}{7}},
  \bibinfo{pages}{011026} (\bibinfo{year}{2017}).

\bibitem[{Vaj()}]{Vajna17}
\bibinfo{howpublished}{S.~Vajna, K.~Klobas, T.~Prosen, and A.~Polkovnikov,
  arXiv:1707.08987 (unpublished).}

\bibitem[{\citenamefont{von Keyserlingk and Sondhi}(2016)}]{vonKeyserlingk16a}
\bibinfo{author}{\bibfnamefont{C.~W.} \bibnamefont{von Keyserlingk}}
  \bibnamefont{and} \bibinfo{author}{\bibfnamefont{S.~L.}
  \bibnamefont{Sondhi}}, \bibinfo{journal}{Phys. Rev. B}
  \textbf{\bibinfo{volume}{93}}, \bibinfo{pages}{245145}
  (\bibinfo{year}{2016}).

\bibitem[{\citenamefont{Else and Nayak}(2016)}]{Else16a}
\bibinfo{author}{\bibfnamefont{D.~V.} \bibnamefont{Else}} \bibnamefont{and}
  \bibinfo{author}{\bibfnamefont{C.}~\bibnamefont{Nayak}},
  \bibinfo{journal}{Phys. Rev. B} \textbf{\bibinfo{volume}{93}},
  \bibinfo{pages}{201103} (\bibinfo{year}{2016}).

\bibitem[{\citenamefont{Roy and Harper}(2016)}]{Roy16}
\bibinfo{author}{\bibfnamefont{R.}~\bibnamefont{Roy}} \bibnamefont{and}
  \bibinfo{author}{\bibfnamefont{F.}~\bibnamefont{Harper}},
  \bibinfo{journal}{Phys. Rev. B} \textbf{\bibinfo{volume}{94}},
  \bibinfo{pages}{125105} (\bibinfo{year}{2016}).

\bibitem[{\citenamefont{Potter et~al.}(2016)\citenamefont{Potter, Morimoto, and
  Vishwanath}}]{Potter16}
\bibinfo{author}{\bibfnamefont{A.~C.} \bibnamefont{Potter}},
  \bibinfo{author}{\bibfnamefont{T.}~\bibnamefont{Morimoto}}, \bibnamefont{and}
  \bibinfo{author}{\bibfnamefont{A.}~\bibnamefont{Vishwanath}},
  \bibinfo{journal}{Phys. Rev. X} \textbf{\bibinfo{volume}{6}},
  \bibinfo{pages}{041001} (\bibinfo{year}{2016}).

\bibitem[{\citenamefont{Potirniche et~al.}(2017)\citenamefont{Potirniche,
  Potter, Schleier-Smith, Vishwanath, and Yao}}]{Potirniche17}
\bibinfo{author}{\bibfnamefont{I.-D.} \bibnamefont{Potirniche}},
  \bibinfo{author}{\bibfnamefont{A.~C.} \bibnamefont{Potter}},
  \bibinfo{author}{\bibfnamefont{M.}~\bibnamefont{Schleier-Smith}},
  \bibinfo{author}{\bibfnamefont{A.}~\bibnamefont{Vishwanath}},
  \bibnamefont{and} \bibinfo{author}{\bibfnamefont{N.~Y.} \bibnamefont{Yao}},
  \bibinfo{journal}{Phys. Rev. Lett.} \textbf{\bibinfo{volume}{119}},
  \bibinfo{pages}{123601} (\bibinfo{year}{2017}).

\bibitem[{\citenamefont{von Keyserlingk et~al.}(2016)\citenamefont{von
  Keyserlingk, Khemani, and Sondhi}}]{vonKeyserlingk16b}
\bibinfo{author}{\bibfnamefont{C.~W.} \bibnamefont{von Keyserlingk}},
  \bibinfo{author}{\bibfnamefont{V.}~\bibnamefont{Khemani}}, \bibnamefont{and}
  \bibinfo{author}{\bibfnamefont{S.~L.} \bibnamefont{Sondhi}},
  \bibinfo{journal}{Phys. Rev. B} \textbf{\bibinfo{volume}{94}},
  \bibinfo{pages}{085112} (\bibinfo{year}{2016}).

\bibitem[{\citenamefont{Else et~al.}(2016)\citenamefont{Else, Bauer, and
  Nayak}}]{Else16b}
\bibinfo{author}{\bibfnamefont{D.~V.} \bibnamefont{Else}},
  \bibinfo{author}{\bibfnamefont{B.}~\bibnamefont{Bauer}}, \bibnamefont{and}
  \bibinfo{author}{\bibfnamefont{C.}~\bibnamefont{Nayak}},
  \bibinfo{journal}{Phys. Rev. Lett.} \textbf{\bibinfo{volume}{117}},
  \bibinfo{pages}{090402} (\bibinfo{year}{2016}).

\bibitem[{\citenamefont{Yao et~al.}(2017)\citenamefont{Yao, Potter, Potirniche,
  and Vishwanath}}]{Yao17}
\bibinfo{author}{\bibfnamefont{N.~Y.} \bibnamefont{Yao}},
  \bibinfo{author}{\bibfnamefont{A.~C.} \bibnamefont{Potter}},
  \bibinfo{author}{\bibfnamefont{I.-D.} \bibnamefont{Potirniche}},
  \bibnamefont{and}
  \bibinfo{author}{\bibfnamefont{A.}~\bibnamefont{Vishwanath}},
  \bibinfo{journal}{Phys. Rev. Lett.} \textbf{\bibinfo{volume}{118}},
  \bibinfo{pages}{030401} (\bibinfo{year}{2017}).

\bibitem[{\citenamefont{Zhang et~al.}(2017)\citenamefont{Zhang, Hess,
  Kyprianidis, Becker, Lee, Smith, Pagano, Potirniche, Potter, Vishwanath
  et~al.}}]{Zhang17}
\bibinfo{author}{\bibfnamefont{J.}~\bibnamefont{Zhang}},
  \bibinfo{author}{\bibfnamefont{P.}~\bibnamefont{Hess}},
  \bibinfo{author}{\bibfnamefont{A.}~\bibnamefont{Kyprianidis}},
  \bibinfo{author}{\bibfnamefont{P.}~\bibnamefont{Becker}},
  \bibinfo{author}{\bibfnamefont{A.}~\bibnamefont{Lee}},
  \bibinfo{author}{\bibfnamefont{J.}~\bibnamefont{Smith}},
  \bibinfo{author}{\bibfnamefont{G.}~\bibnamefont{Pagano}},
  \bibinfo{author}{\bibfnamefont{I.}~\bibnamefont{Potirniche}},
  \bibinfo{author}{\bibfnamefont{A.}~\bibnamefont{Potter}},
  \bibinfo{author}{\bibfnamefont{A.}~\bibnamefont{Vishwanath}},
  \bibnamefont{et~al.}, \bibinfo{journal}{Nature}
  \textbf{\bibinfo{volume}{543}}, \bibinfo{pages}{217} (\bibinfo{year}{2017}).

\bibitem[{\citenamefont{Choi et~al.}(2017)\citenamefont{Choi, Choi, Landig,
  Kucsko, Zhou, Isoya, Jelezko, Onoda, Sumiya, Khemani et~al.}}]{Choi17}
\bibinfo{author}{\bibfnamefont{S.}~\bibnamefont{Choi}},
  \bibinfo{author}{\bibfnamefont{J.}~\bibnamefont{Choi}},
  \bibinfo{author}{\bibfnamefont{R.}~\bibnamefont{Landig}},
  \bibinfo{author}{\bibfnamefont{G.}~\bibnamefont{Kucsko}},
  \bibinfo{author}{\bibfnamefont{H.}~\bibnamefont{Zhou}},
  \bibinfo{author}{\bibfnamefont{J.}~\bibnamefont{Isoya}},
  \bibinfo{author}{\bibfnamefont{F.}~\bibnamefont{Jelezko}},
  \bibinfo{author}{\bibfnamefont{S.}~\bibnamefont{Onoda}},
  \bibinfo{author}{\bibfnamefont{H.}~\bibnamefont{Sumiya}},
  \bibinfo{author}{\bibfnamefont{V.}~\bibnamefont{Khemani}},
  \bibnamefont{et~al.}, \bibinfo{journal}{Nature}
  \textbf{\bibinfo{volume}{543}}, \bibinfo{pages}{221} (\bibinfo{year}{2017}).

\bibitem[{\citenamefont{Asb\'oth et~al.}(2014)\citenamefont{Asb\'oth,
  Tarasinski, and Delplace}}]{Asboth14}
\bibinfo{author}{\bibfnamefont{J.~K.} \bibnamefont{Asb\'oth}},
  \bibinfo{author}{\bibfnamefont{B.}~\bibnamefont{Tarasinski}},
  \bibnamefont{and} \bibinfo{author}{\bibfnamefont{P.}~\bibnamefont{Delplace}},
  \bibinfo{journal}{Phys. Rev. B} \textbf{\bibinfo{volume}{90}},
  \bibinfo{pages}{125143} (\bibinfo{year}{2014}).

\bibitem[{\citenamefont{Jackiw and Rebbi}(1976)}]{JackiwRebbi}
\bibinfo{author}{\bibfnamefont{R.}~\bibnamefont{Jackiw}} \bibnamefont{and}
  \bibinfo{author}{\bibfnamefont{C.}~\bibnamefont{Rebbi}},
  \bibinfo{journal}{Phys. Rev. D} \textbf{\bibinfo{volume}{13}},
  \bibinfo{pages}{3398} (\bibinfo{year}{1976}).

\bibitem[{\citenamefont{Su et~al.}(1979)\citenamefont{Su, Schrieffer, and
  Heeger}}]{SSH}
\bibinfo{author}{\bibfnamefont{W.~P.} \bibnamefont{Su}},
  \bibinfo{author}{\bibfnamefont{J.~R.} \bibnamefont{Schrieffer}},
  \bibnamefont{and} \bibinfo{author}{\bibfnamefont{A.~J.}
  \bibnamefont{Heeger}}, \bibinfo{journal}{Phys. Rev. Lett.}
  \textbf{\bibinfo{volume}{42}}, \bibinfo{pages}{1698} (\bibinfo{year}{1979}).

\bibitem[{\citenamefont{Jackiw and Rossi}(1981)}]{JackiwRossi}
\bibinfo{author}{\bibfnamefont{R.}~\bibnamefont{Jackiw}} \bibnamefont{and}
  \bibinfo{author}{\bibfnamefont{P.}~\bibnamefont{Rossi}},
  \bibinfo{journal}{Nuclear Physics B} \textbf{\bibinfo{volume}{190}},
  \bibinfo{pages}{681 } (\bibinfo{year}{1981}).

\bibitem[{\citenamefont{Read and Green}(2000)}]{ReadGreen}
\bibinfo{author}{\bibfnamefont{N.}~\bibnamefont{Read}} \bibnamefont{and}
  \bibinfo{author}{\bibfnamefont{D.}~\bibnamefont{Green}},
  \bibinfo{journal}{Phys. Rev. B} \textbf{\bibinfo{volume}{61}},
  \bibinfo{pages}{10267} (\bibinfo{year}{2000}).

\bibitem[{\citenamefont{Teo and Kane}(2010)}]{TeoKane}
\bibinfo{author}{\bibfnamefont{J.~C.~Y.} \bibnamefont{Teo}} \bibnamefont{and}
  \bibinfo{author}{\bibfnamefont{C.~L.} \bibnamefont{Kane}},
  \bibinfo{journal}{Phys. Rev. B} \textbf{\bibinfo{volume}{82}},
  \bibinfo{pages}{115120} (\bibinfo{year}{2010}).

\bibitem[{Ati()}]{AtiyahSinger}
\bibinfo{howpublished}{M. F. Atiyah and I. M. Singer, Bull. Amer. Math. Soc.
  \textbf{69}, 422 (1963).}

\bibitem[{\citenamefont{Weinberg}(1981)}]{Weinberg}
\bibinfo{author}{\bibfnamefont{E.~J.} \bibnamefont{Weinberg}},
  \bibinfo{journal}{Phys. Rev. D} \textbf{\bibinfo{volume}{24}},
  \bibinfo{pages}{2669} (\bibinfo{year}{1981}).

\bibitem[{\citenamefont{Nicolai}(1976)}]{Nicolai76}
\bibinfo{author}{\bibfnamefont{H.}~\bibnamefont{Nicolai}},
  \bibinfo{journal}{Journal of Physics A: Mathematical and General}
  \textbf{\bibinfo{volume}{9}}, \bibinfo{pages}{1497} (\bibinfo{year}{1976}).

\bibitem[{\citenamefont{Witten}(1981)}]{Witten81}
\bibinfo{author}{\bibfnamefont{E.}~\bibnamefont{Witten}},
  \bibinfo{journal}{Nucl. Phys. B} \textbf{\bibinfo{volume}{188}},
  \bibinfo{pages}{513 } (\bibinfo{year}{1981}).

\bibitem[{\citenamefont{Witten}(1982)}]{Witten82}
\bibinfo{author}{\bibfnamefont{E.}~\bibnamefont{Witten}},
  \bibinfo{journal}{Nucl. Phys. B} \textbf{\bibinfo{volume}{202}},
  \bibinfo{pages}{253 } (\bibinfo{year}{1982}).

\bibitem[{\citenamefont{Fendley and Schoutens}(2005)}]{Fendley05}
\bibinfo{author}{\bibfnamefont{P.}~\bibnamefont{Fendley}} \bibnamefont{and}
  \bibinfo{author}{\bibfnamefont{K.}~\bibnamefont{Schoutens}},
  \bibinfo{journal}{Phys. Rev. Lett.} \textbf{\bibinfo{volume}{95}},
  \bibinfo{pages}{046403} (\bibinfo{year}{2005}).

\bibitem[{\citenamefont{van Eerten}(2005)}]{vanEerten05}
\bibinfo{author}{\bibfnamefont{H.}~\bibnamefont{van Eerten}},
  \bibinfo{journal}{J.~Math.~Phys.} \textbf{\bibinfo{volume}{46}},
  \bibinfo{pages}{123302} (\bibinfo{year}{2005}).

\bibitem[{Kat()}]{Katsura17}
\bibinfo{howpublished}{H. Katsura, H. Moriya, and Y. Nakayama, arXiv:1710.04385
  (unpublished).}

\bibitem[{SM()}]{SM}
\bibinfo{howpublished}{See Supplementary Material.}

\bibitem[{\citenamefont{Pal and Huse}(2010)}]{Pal10}
\bibinfo{author}{\bibfnamefont{A.}~\bibnamefont{Pal}} \bibnamefont{and}
  \bibinfo{author}{\bibfnamefont{D.~A.} \bibnamefont{Huse}},
  \bibinfo{journal}{Phys. Rev. B} \textbf{\bibinfo{volume}{82}},
  \bibinfo{pages}{174411} (\bibinfo{year}{2010}).

\bibitem[{\citenamefont{Hsieh et~al.}(2016)\citenamefont{Hsieh, Hal\'asz, and
  Grover}}]{Hsieh16}
\bibinfo{author}{\bibfnamefont{T.~H.} \bibnamefont{Hsieh}},
  \bibinfo{author}{\bibfnamefont{G.~B.} \bibnamefont{Hal\'asz}},
  \bibnamefont{and} \bibinfo{author}{\bibfnamefont{T.}~\bibnamefont{Grover}},
  \bibinfo{journal}{Phys, Rev. Lett.} \textbf{\bibinfo{volume}{117}},
  \bibinfo{pages}{166802} (\bibinfo{year}{2016}).

\end{thebibliography}

\appendix
\begin{widetext}
\begin{center}
\textbf{\large Supplementary Material for ``Floquet Supersymmetry"}
\end{center}

\section{Appendix A: Degeneracy as Intersection of Subspaces}

\setcounter{equation}{0}
\renewcommand{\theequation}{A\arabic{equation}}

Here we provide an alternative proof for the $2^{L/2}$ degeneracy and show that it arises from the intersection of two large subspaces of the Hilbert space.  For convenience, we assume $L$ is a multiple of $4$; it is straightforward to extend the proof below to other even $L$.  First observe that the operator $U^{\,}_{ZZ}$ has eigenvalues $\pm 1$.  This is because in any product state in the $Z$ basis, $U^{\,}_{ZZ}$ produces a phase $e^{-\mathrm{i}\pi/4}$ for every domain wall and $e^{\mathrm i\pi/4}$ otherwise.  Since, the total number of domain walls and non-domain walls must add to $L$, a multiple of $4$, the total phase produced is $\pm 1$.  

Define $\mathcal S$ to be the span of all product states in the $Z$ basis on which the two operators $U^{\,}_{ZZ}$ and $Q \equiv (-1)^{L/4}  \prod^{L}_{i=1}Z^{\,}_{i}$ have identical action:
\begin{align}
\mathcal S
=
\text{span}
\left\{\ket{z}
=
\ket{z^{\,}_{1}\dots z^{\,}_{L}}
\mid
U^{\,}_{ZZ}\!\ket{z}
=
Q\!\ket{z} \right\}.
\end{align}
Both $U^{\,}_{ZZ}$ and $Q$ have eigenvalues $\pm 1$, so to determine the dimension of $\mathcal S$, we need to count how many eigenvalues of $U^{\,}_{ZZ} Q$ are 1.  This is achieved by evaluating 
\begin{align}
\begin{split}
\text{tr}(U^{\,}_{ZZ} Q)
&= (-1)^{L/4} \left(\frac{1}{\sqrt{2}}\right)^L \text{tr} \prod^{L}_{i=1} (1+iZ^{\,}_i Z^{\,}_{i+1})  \prod^{L}_{i=1}Z^{\,}_{i}\\
&= 2^{L/2 + 1},
\end{split}
\end{align}  
where we have used the fact that any operator string involving a Pauli operator is traceless.  Since the total number of eigenvalues is $2^L$, we deduce that the number of eigenvalues 1, and hence the dimension of $\mathcal S$, is $2^{L-1} + 2^{L/2}$.  

Consider the projection $P^{\,}_{\mathcal S}$ into the subspace $\mathcal S$.  By appending this projection to the time evolution $U^{\,}_{\rm F} = U^{\,}_{ZZ} U^{\,}_H$, we can trade $U^{\,}_{ZZ}$ for the simpler $Q$: 
\begin{align}
P^{\,}_{\mathcal S}\, U^{\,}_{\rm F}
=
P^{\,}_{\mathcal S}\, Q\, U^{\,}_H
=
P^{\,}_{\mathcal S}\, \widetilde{U}^{\,}_{\rm F}, \label{eq: projected}
\end{align}
where
\begin{align}
\widetilde{U}^{\,}_{\rm F}
=Q\, U^{\,}_H
=
U_H^{1/2 \dagger}
Q
U_H^{1/2}.
\end{align}
For the last equality, we have used the fact that all interactions in the argument of 
\begin{align}
U^{\,}_H = \exp\[-\frac{\mathrm{i}}{2}\(\sum^{L}_{i=1}h^{\,}_{i}\, X^{\,}_{i}+g\, H^{\,}_{\rm int}\)\],
\end{align}
are assumed to anticommute with $Q$. Hence, $\widetilde{U}^{\,}_{\rm F}$ is simply a rotated $Q$ operator and thus has two $2^{L-1}$-dimensional eigenspaces $\widetilde{\mathcal H}^{\,}_{\pm}$ with eigenvalues $\pm1$.

Returning to \eqref{eq: projected}, we see that any state in $\mathcal S\cap \widetilde{\mathcal H}^{\,}_{\pm}$ will be an eigenstate of $U^{\,}_{\rm F}$ with eigenvalue $\pm1$.  Thus finding a lower bound on the dimension of $\mathcal S\cap \widetilde{\mathcal H}^{\,}_{\pm}$ will provide the desired lower bound on the $0,\pi$ degeneracies.  Finding this lower bound on the intersection is a pigeonhole-principle-type argument, which we make precise below. 

We decompose
\begin{align}
\mathcal S = (\mathcal S\cap\widetilde{\mathcal H}^{\,}_{+})\oplus \mathcal S^{\prime}_{+},
\end{align}
where 
$\mathcal S^{\prime}_{+}$ is the orthogonal complement of $\mathcal S\cap\widetilde{\mathcal H}^{\,}_{+}$
 in $\mathcal S$. Now, define $P^{\,}_{-}$, the projector into $\widetilde{\mathcal H}^{\,}_{-}$. For any state
 $\ket{v}\in\mathcal S^{\prime}_{+}$, $P^{\,}_{-}\ket{v}\neq0$; if this were not true, then it would follow that
 $\ket{v}\in\mathcal S\cap\widetilde{\mathcal H}^{\,}_{+}$, which is a contradiction.
 This means that for any $\ket{v^{\,}_{1}},\ket{v^{\,}_{2}}\in\mathcal S^{\prime}_{+}$,
 \begin{align}
 P^{\,}_{-}\ket{v^{\,}_{1}}=P^{\,}_{-}\ket{v^{\,}_{2}}\implies\ket{v^{\,}_{1}}=\ket{v^{\,}_{2}},
 \end{align}
 since otherwise the state $\ket{v^{\,}_{1}}-\ket{v^{\,}_{2}}\in\mathcal S^{\prime}_{+}$ would be
 annihilated by $P^{\,}_{-}$.
 Hence $P^{\,}_{-}$ is injective as a function
 from $\mathcal S^{\prime}_{+}$ into $\widetilde{\mathcal H}^{\,}_{-}$. It follows that
 $\text{dim}\ \mathcal S^{\prime}_{+}\leq \text{dim}\ \widetilde{\mathcal H}^{\,}_{-}=2^{L-1}$.
Since $\text{dim}\ \mathcal S=2^{L-1}+2^{L/2}$,
we conclude that $\text{dim}\, \mathcal S\cap\widetilde{\mathcal H}^{\,}_{+}\geq 2^{L/2}$.
A similar argument can be applied to lower-bound
$\text{dim}\, \mathcal S\cap\widetilde{\mathcal H}^{\,}_{-}$.  Thus, we have proved that
the eigenspaces at quasienergy $0$ and $\pi$ are each at least $2^{L/2}$-fold degenerate.

This proof provides some intuition on the nature of the degenerate states: since they arise from the intersection of two extensive subspaces, one expects them to be generically highly entangled.   

\section{Appendix B: Dynamics of Restricted Floquet Evolution}

Denote the union of the degenerate subspaces at quasienergies $0$ and $\pi$ by $D$, and denote the projector onto $D$ by $P^{\,}_D$.  From the above appendix, the Floquet evolution restricted to $D$ is given by $\tU^{\,}_{\rm F} P^{\,}_D$.  Recall that, up to a sign, $\tU^{\,}_{\rm F}$ is $\prod^{L}_{i=1}Z^{\,}_{i}$ rotated by a unitary $U^{1/2}_H$.  It then follows that $U^{1/2 \dagger}_H X^{\,}_i\, U^{1/2}_H$ flips sign once per period of the projected Floquet evolution.  In the simplest model (7), $U^{\,}_H = U^{\,}_X$ commutes with $X^{\,}_i$, so the on-site operator $X^{\,}_i$ flips sign every period of the projected Floquet evolution.

\section{Appendix C: Effects of Weak Time-Reflection Symmetry Breaking}

\setcounter{equation}{0}
\setcounter{figure}{0}
\renewcommand{\theequation}{C\arabic{equation}}
\renewcommand{\thefigure}{C\arabic{figure}}

The phenomena investigated in the main text rely crucially on the existence of a time-reflection operator $R$ that maps the Floquet operator $U^{\,}_{\rm F}$ to its adjoint.  In realistic quantum simulators, such a scenario may only be true to a finite degree of accuracy.  In the Floquet drive (7), such deviations can arise when the desired angle $\pi/4$ entering $U^{\,}_{ZZ}$ differs slightly from the actual angle realized during the experimental pulse sequence.   This raises the question of whether the characteristic subharmonic response visible in Fig.~2 remains observable in the presence of such pulse imperfections. 

To investigate this question, we consider the evolution out to a time $t^{\,}_{\rm f}$ generated by the modified evolution operator
\begin{subequations}
\label{eq: modified evolution}
\begin{align}
\label{eq: modified evolution a}
U^{\,}_{\{\epsilon\}}(t^{\,}_{\rm f})=\prod^{t^{\,}_{\rm f}}_{t=0}U^{\,}_{ZZ}(\epsilon^{\,}_{t})\, U^{\,}_{H},
\end{align}
where
\begin{align}
\label{eq: modified evolution b}
U^{\,}_{ZZ}(\epsilon^{\,}_{t}) \equiv \text{exp}\left[\mathrm{i}\frac{\pi}{4}\, (1+\epsilon^{\,}_t)\sum^{L}_{i=1}Z^{\,}_{i}Z^{\,}_{i+1}\right]. 
\end{align}
\end{subequations}
The pulse error $\epsilon^{\,}_{t}$ above can be chosen to vary with the time step $t$, to model cases where the experimental apparatus is subject to some random noise or parameter drift as a function of time.  We single out two cases of interest: ``coherent errors," where $\epsilon^{\,}_{t}\equiv \epsilon$ is time-independent but nonzero, and ``incoherent errors," where $\epsilon^{\,}_{t}$ is chosen randomly from a uniform distribution with mean zero.  The former case preserves the discrete time translation symmetry of the Floquet problem, while the latter case is perhaps more experimentally relevant, as it models a scenario where the target pulse sequence is imperfectly realized in a slightly different manner each time step.  One figure of merit to quantify the persistence or absence of subharmonic response is the staggered moving average of the $X$-basis magnetization,
\begin{align}
\label{eq: staggered magnetization}
\overline{\langle\tilde{M}^{\,}_{X}\rangle}_{t}
=
\frac{1}{t}\sum^{t}_{t'=0}(-1)^t\, \langle M^{\,}_{X}\rangle.
\end{align}
When $\epsilon_t\equiv 0$, the disorder average of Eq.~\eqref{eq: staggered magnetization} saturates to a nonvanishing constant value as a function of time, indicating the presence of persistent period-2 oscillations.  When $\epsilon_t$ is finite, we expect the period-2 oscillations to decay on a time scale that depends on the choice of $\{\epsilon^{\,}_t\}$.

\begin{figure}[t]
(a)\includegraphics[width=0.45\textwidth]{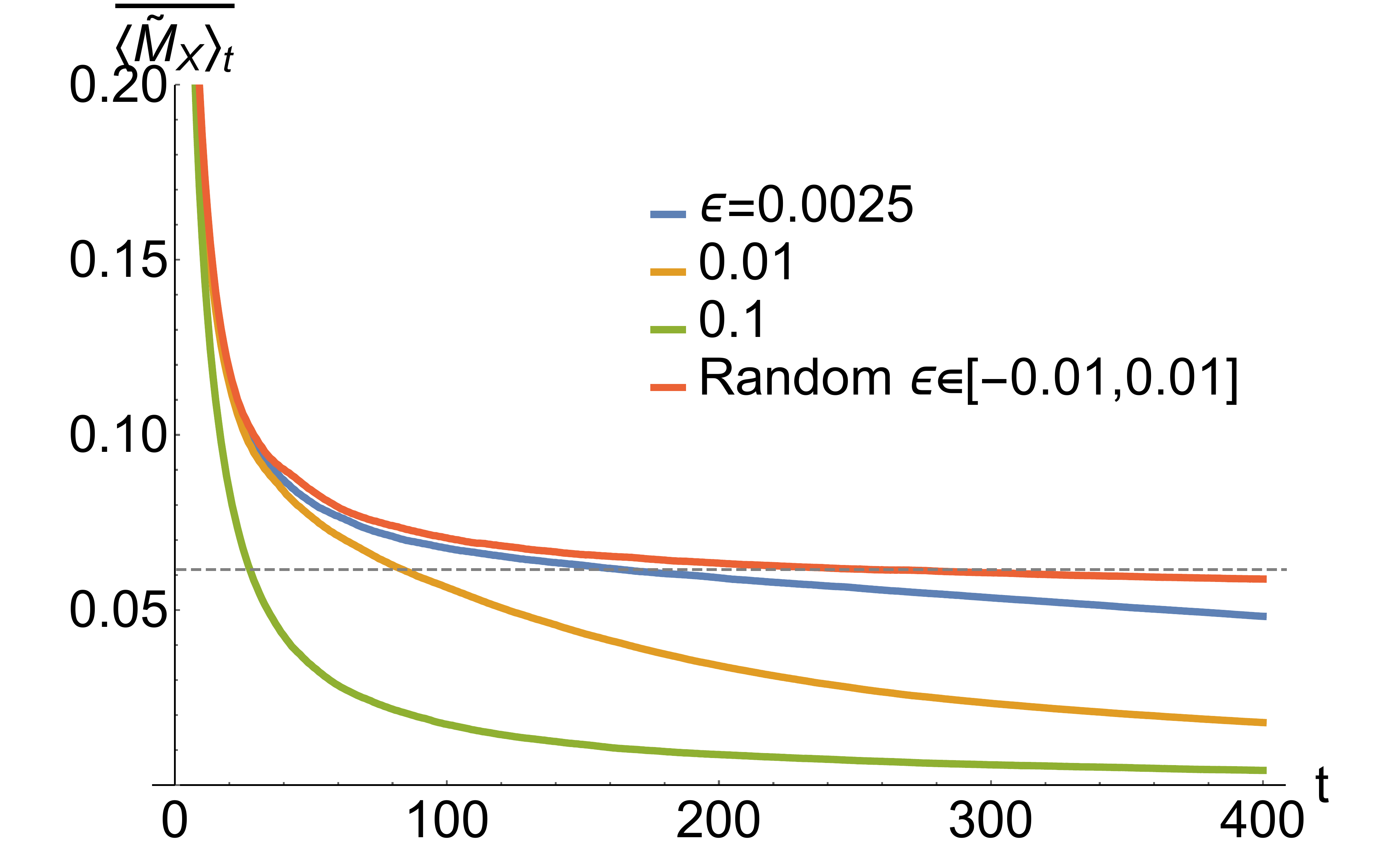}
(b)\includegraphics[width=0.45\textwidth]{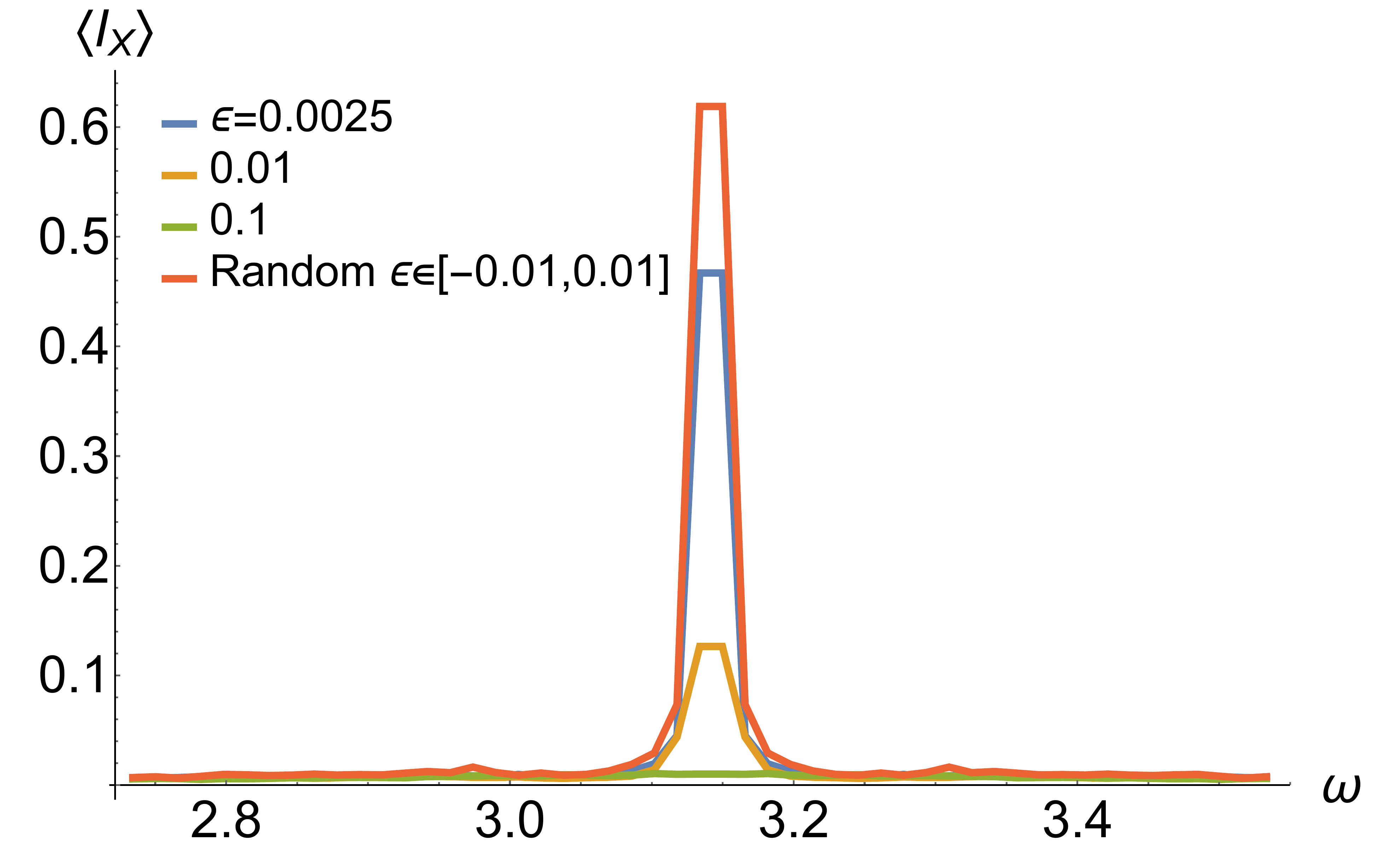}
\caption{
(Color online)
Investigating subharmonic response in the presence of time-reflection-breaking pulse errors.  (a) Staggered moving average of the magnetization, Eq.~\eqref{eq: staggered magnetization},
under the modified dynamics \eqref{eq: modified evolution}.  Coherent errors (blue, orange, and green) lead to a decay of the subharmonic response with a time-scale of order $1/\epsilon$.  Incoherent errors (red) significantly enhance the lifetime of the subharmonic response relative to the case of coherent errors of similar magnitude.  (b) Power spectrum of the magnetization illustrating the suppression and broadening of the $\pi$-peak with increasing error strength, as well as the enhancement of the peak height in the case of incoherent errors.
        }
\label{fig: pulse imperfection data}
\end{figure}

Our numerical results for the figure of merit \eqref{eq: staggered magnetization} under the modified dynamics \eqref{eq: modified evolution} for both the coherent and incoherent cases is shown in Fig.~\ref{fig: pulse imperfection data}(a).  The data are computed at $L=8$ and $g=1$ using the same initial state as in Fig.~2, and are averaged over 100 disorder realizations.

In the case of coherent errors, the period-2 oscillations decay on a time scale of order $1/\epsilon$.  For example, for $\epsilon=0.01$, indicating a consistent overrotation by $1\%$ of the desired angle $\pi/4$ at each time step, appreciable period-doubled oscillations remain out to a time of order 100 driving periods.  The corresponding power spectra $\langle I^{\,}_{X}(\omega)\rangle$ shown in Fig.~\ref{fig: pulse imperfection data}(b) indicate that coherent errors suppress the height of the peak at $\omega=\pi$ that is characteristic of the period-2 oscillations, in addition to inducing some broadening relative to the error-free case. Nevertheless, a clear peak at $\omega=\pi$ is visible for sufficiently small pulse imperfections. 

To simulate the incoherent case, we chose $\epsilon^{\,}_{t}$ uniformly and at random from the interval $[-0.01,0.01]$, corresponding to a hypothetical $1\%$ error in the desired pulse that is uncorrelated between successive time steps.  In this case, the lifetime of the period-2 oscillations is significantly enhanced relative to the case of a fixed $1\%$ error per time step.  This is due to the fact that these incoherent errors essentially ``self-average" during the course of the evolution, and the chosen distribution of errors has mean zero.

These results suggest that, for an experiment at fixed system size that is capable of maintaining quantum coherence out to a time $t^{\,}_{\rm f}$, there is a threshold in the severity of time-reflection symmetry breaking below which the system exhibits subharmonic response at all experimentally accessible times.

\end{widetext}

\end{document}